\documentclass[prl,
reprint,
superscriptaddress,
showpacs,preprintnumbers,
nofootinbib,
amsmath,amssymb,
aps,
]{revtex4-1}

\usepackage{graphicx}
\usepackage{dcolumn}
\usepackage{bm}
\usepackage{color}

\newcommand{\beq}{\begin{equation}}
\newcommand{\eeq}{\end{equation}}
\newcommand{\bea}{\begin{eqnarray}}
\newcommand{\eea}{\end{eqnarray}}
\def\dif{\mathrm{d}}
\def\le{\left}
\def\ri{\right}
\def\F{\Phi}
\def\pa{\partial}

\newcommand*{\mcO}{\mathcal{O}}

\newcommand*{\half}{\frac{1}{2}}

\newcommand*{\hc}{\dagger}
\newcommand*{\Gm}{\Gamma}
\newcommand*{\w}{\omega}
\newcommand{\godo}{G_{\mathcal{O}^{\dagger}\mathcal{O}}}
\newcommand{\good}{G_{\mathcal{O}\mathcal{O}^{\dagger}}}
\newcommand{\goo}{G_{\mathcal{O}\mathcal{O}}}

\newcommand{\rodo}{\rho_{\mathcal{O}^{\dagger}\mathcal{O}}}

\begin{document}

\preprint{FPAUO-16/15, OUTP-16-26P, SISSA 60/2016/FISI}

\title{Holographic Kondo and Fano Resonances}

\author{Johanna Erdmenger}
\email{erdmenger@physik.uni-wuerzburg.de}
\affiliation{Institut f\"ur Theoretische Physik und Astrophysik, Julius-Maximilians-Universit\"at W\"urzburg, Am Hubland, D-97074 W\"urzburg, Germany.\\ Max-Planck-Institut f\"ur Physik (Werner-Heisenberg-Institut),\\
F\"ohringer Ring 6, D-80805 Munich, Germany.}

\author{Carlos Hoyos}
\email{hoyoscarlos@uniovi.es}
\affiliation{Department of Physics, Universidad de Oviedo, Avda.~Calvo Sotelo 18, 33007, Oviedo, Spain.}

\author{Andy O'Bannon}
\email{a.obannon@soton.ac.uk}
\affiliation{STAG Research Centre, Physics and Astronomy, University of Southampton, Southampton SO17 1BJ, U.~K.}

\author{Ioannis~Papadimitriou}
\email{ioannis.papadimitriou@sissa.it}
\affiliation{SISSA and INFN - Sezione di Trieste, Via Bonomea 265, I 34136 Trieste, Italy.}

\author{Jonas Probst}
\email{Jonas.Probst@physics.ox.ac.uk}
\affiliation{Rudolf Peierls Centre for Theoretical Physics, University of Oxford, 1 Keble Road, Oxford OX1~3NP, U.~K.}

\author{Jackson M. S. Wu}
\email{jknwgm13@gmail.com}
\affiliation{Department of Physics and Astronomy, University of Alabama, Tuscaloosa, AL 35487, USA.}

\date{\today}

\begin{abstract}

We use holography to study a $(1+1)$-dimensional Conformal Field Theory (CFT) coupled to an impurity. The CFT is an $SU(N)$ gauge theory at large $N$, with strong gauge interactions. The impurity is an $SU(N)$ spin. We trigger an impurity Renormalization Group (RG) flow via a Kondo coupling. The Kondo effect occurs only below the critical temperature of a large-$N$ mean-field transition. We show that at all temperatures $T$, impurity spectral functions exhibit a Fano resonance, which in the low-$T$ phase is a large-$N$ manifestation of the Kondo resonance. We thus provide an example in which the Kondo resonance survives strong correlations, and uncover a novel mechanism for generating Fano resonances, via RG flows between $(0+1)$-dimensional fixed points.

\end{abstract}

\maketitle

\vspace{-0.4cm}
\section{Introduction}
\vspace{-0.3cm}

The Kondo effect is the screening of an impurity spin by a Landau Fermi Liquid (LFL) at low $T$~\cite{PTP.32.37,Hewson:1993}. A variety of techniques, such as Wilson's RG, large-$N$, CFT, and more~\cite{doi:10.1080/000187398243500}, have captured many characteristic Kondo phenomena. Nevertheless, many questions resist solution, for example about inter-impurity interactions, subsystem Entanglement Entropy (EE), non-equilibrium processes like quantum quenches, and more.

In particular, what happens when the LFL is replaced with strongly correlated electrons? For example, how does the Kondo effect change in a Luttinger liquid~\cite{PhysRevLett.69.3378,PhysRevLett.72.892,PhysRevLett.75.300,PhysRevB.53.3211,2005JPSJ...74...73F} or the Hubbard model~\cite{Fulde1993,1994PhRvB..50.1345S}? In the latter case, experiments reveal dramatic effects of strong correlations, such as enhancement of the Kondo temperature, $T_K$. On the theory side, although special tools like bosonization~\cite{PhysRevLett.69.3378,PhysRevLett.72.892,PhysRevLett.75.300,PhysRevB.53.3211,2005JPSJ...74...73F} and uncontrolled mean-field approximations~\cite{Fulde1993,PhysRevB.52.9514,PhysRevB.52.15966,1997cond.mat..1137T,1997PhRvB..56.6559S,1998PhRvB..57.7773D,doi:10.1142/S0217984999001044,2000PhRvL..84.4417H} have provided insight, in general, reliable techniques do not yet exist to answer questions about Kondo phenomena in strongly-correlated systems.

To address all of the above, we have developed an alternative Kondo model, based on holographic duality~\cite{Erdmenger:2013dpa,O'Bannon:2015gwa,Erdmenger:2015spo,Erdmenger:2015xpq}. Our model replaces the LFL by a $(1+1)$-dimensional CFT in which spin $SU(2)$ is replaced by \textit{gauged} $SU(N)$, with large $N$ and strong gauge interactions. Our model has already revealed novel strong-coupling phenomena in RG~\cite{Erdmenger:2013dpa,O'Bannon:2015gwa}, inter-impurity interactions~\cite{O'Bannon:2015gwa} and EE~\cite{Erdmenger:2015spo}.

Here we initiate the study of non-equilibrium phenomena in our model: we compute linear response (Green's) functions of a charged bosonic impurity operator, $\mathcal{O}$, in our model. We have two main results.

First, we find a large-$N$ manifestation of the \textit{Kondo resonance}~\cite{Hewson:1993,Phillips2012,Coleman2015}, a signature of the Kondo effect. As expected, our Kondo resonance appears only at $T$ below the critical temperature $T_c$ of a mean-field transition that is common to large-$N$ Kondo models~\cite{0022-3719-19-17-017,PhysRevB.35.5072,2003PhRvL..90u6403S,2004PhRvB..69c5111S,Coleman2015}: $\langle \mathcal{O} \rangle$ becomes non-zero when $T \leq T_c$. We thus prove unequivocally that our holographic model realizes a genuine Kondo effect, as opposed to some other impurity physics, and furthermore show that a large-$N$ Kondo resonance can survive strong correlations essentially intact.

Second, at all $T$, $\mathcal{O}$'s spectral function exhibits a \textit{Fano resonance}~\cite{PhysRev.124.1866,RevModPhys.82.2257}, which occurs when a Lorentzian resonance is immersed in a continuum of states (in energy). A Fano resonance is characterized not only by its position, width, and height, like a Lorentzian, but also by an \textit{asymmetry parameter}, $q$, which measures the relative strength of resonant versus non-resonant scattering. Our $q$ increases as $T\to T_c^+$. When $T \leq T_c$, the Fano line-shape arises from our Kondo resonance, which must be \textit{anti-symmetric} due to Particle-Hole Symmetry (PHS), and hence has the special value $q=1$.

Although Fano resonances have been observed in many impurity systems in one spatial dimension~\cite{Madhavan567,PhysRevB.64.165412,2000PhRvB..62.2188G,RevModPhys.82.2257}, ours arise from a qualitatively different mechanism.
For instance, in side-coupled quantum dots (QDs)~\cite{PhysRevB.64.165412,2000PhRvB..62.2188G,RevModPhys.82.2257} the Lorentzian resonances are the discrete states on the QD, and the continuum comes from electronic scattering states in the leads. Coupling the two, for example by a Kondo coupling, can then produce Fano resonances.

Our model also has an impurity coupled to a continuum in one spatial dimension, \textit{i.e.}~the CFT. However, our model has \textit{two} couplings: the CFT's $SU(N)$ gauge coupling and the Kondo coupling. The spectral function of $\mathcal{O}$ inherits $(0+1)$-dimensional scale invariance from the former, and so exhibits a continuum of states, in contrast to a QD's discrete states. The Kondo coupling then triggers an RG flow from that $(0+1)$-dimensional fixed point, and creates a resonance that cannot escape the continuum, hence producing a Fano line-shape.

To our knowledge, such a mechanism for producing Fano resonances is novel, and moreover is easy to generalize to any RG flow between $(0+1)$-dimensional fixed points, as follows. Scale invariance implies that any spectral function will be a featureless continuum, which in $(0+1)$ dimensions means a power law (or logarithm) in frequency. A relevant deformation can then explicitly break scale invariance, trigger an RG flow to an IR fixed point---in which case we expect the continuum to survive---and may also produce resonances. In higher dimensions, the resonances would not have to be within the continuum, for example the two could be separated in momentum space. However, in $(0+1)$ dimensions the resonances have no place to escape the continuum, and hence must produce Fano line-shapes.

In fact, such a mechanism was at work in some previous cases, such as the large-$N$ Kondo model at sufficiently low $T$~\cite{1998PhRvB..58.3794P}, and holographic duals of $T=0$ charged black branes~\cite{Faulkner:2009wj,Faulkner:2011tm,Sachdev:2015efa}. However, the resulting Fano resonances went unidentified, leaving crucial physics overlooked, namely the relative strength of resonant versus non-resonant scattering, as measured by $q$. Our results not only provide a novel perspective on these cases, but also predict Fano resonances in RG flows between other $(0+1)$-dimensional fixed points, such as Sachdev-Ye-Kitaev fixed points~\cite{Sachdev:1992fk,kitaev,Sachdev:2015efa,Polchinski:2016xgd,Jevicki:2016bwu,Maldacena:2016hyu,Jevicki:2016ito,Witten:2016iux}.

Further results of our model, including details of holographic renormalization useful for holographic impurity models in general, will appear in~\cite{Erdmenger:2016jjg}.

\vspace{-0.5cm}
\section{Holographic Kondo Model}
\vspace{-0.3cm}
We first briefly review some essential features of the CFT and large-$N$ approaches to the Kondo model, and how our model builds upon and extends them.

The CFT approach~\cite{Affleck:1995ge} is based on s-wave reduction of LFL fermions about the impurity, plus linearization of the dispersion relation. In/out-going s-waves become relativistic left/right-moving fermions, $\psi_L$ and $\psi_R$, in the radial direction, $r$. Reflecting $\psi_R$ to $r<0$ and relabeling $\psi_R \to \psi_L$ leads to $\psi_L$ alone on the entire $r$ axis, with impurity at $r=0$. The $\psi_L$ form a $(1+1)$-dimensional chiral CFT with $SU(2)_1 \times U(1)$ spin and charge Kac-Moody currents, respectively. In the Hamiltonian, the Kondo interaction is $\delta(r) g_K S^A J^A$, with coupling constant $g_K$, impurity spin $S^A$, and spin current $J^A$, $A=1,2,3$. An antiferromagnetic coupling, $g_K>0$, is marginally relevant, and triggers an RG flow to an IR chiral CFT characterized by a phase shift of $\psi_L$ and impurity screening~\cite{Affleck:1995ge}.

The large-$N$ approach begins by replacing spin $SU(2) \to SU(N)$, followed by $N \to \infty$ with $\lambda_K \equiv N g_K$ fixed~\cite{RevModPhys.59.845,doi:10.1080/000187398243500,2006cond.mat.12006C,Coleman2015}. We will only consider $S^A$ in totally anti-symmetric $SU(N)$ representations of rank $\mathcal{Q}$, and introduce Abrikosov pseudo-fermions $\chi$ via $S^A=\chi^\dagger T^A\chi$, with $SU(N)$ generators $T^A$, $A = 1,\ldots,N^2-1$. Doing so introduces an auxiliary $U(1)$ acting only on $\chi$, but with charge fixed by projecting onto states with $\chi^{\dagger} \chi = \mathcal{Q}$. At large $N$, $S^A J^A = -\mathcal{O}^\dagger\mathcal{O}/2$ with $\mathcal{O}\equiv\psi_L^\dagger\chi$ \cite{O'Bannon:2015gwa}, which is charged under both the charge and auxiliary $U(1)$'s.

Our holographic model~\cite{Erdmenger:2013dpa} begins by \textit{gauging} $SU(N)$, thus introducing the 't Hooft coupling, $\lambda$. We then add degrees of freedom to make the gauge theory a $(1+1)$-dimensional CFT with sparse operator spectrum when $N$ and $\lambda$ both $\to \infty$, but whose details otherwise are irrelevant. The theory is then holographically dual to Einstein-Hilbert gravity in $(2+1)$-dimensional Anti-de Sitter space, $AdS_3$~\cite{Aharony:1999ti}. The charge $U(1)$ Kac-Moody current is dual to a $U(1)$ Chern-Simons gauge field, $A$, the auxiliary $U(1)$ is dual to a Maxwell field $a$ on an $AdS_2$ defect at $r=0$, and $\mathcal{O}$ is dual to a complex scalar field $\F$ also in $AdS_2$, charged under both $A$ and $a$. As long as the stress-energy tensor is finite, at large $N$ we can neglect back-reaction of $A$, $a$, $\F$ (dual to fundamental fields) on the geometry (dual to adjoint fields). When $T>0$, the bulk metric is thus the BTZ black brane,
\beq
ds^2 = \frac{1}{z^2}\le(h^{-1}(z)\dif z^2 -h(z)\dif t^2+\dif r^2\ri), \nonumber
\eeq
with $h(z)= 1-z^2/z_H^2$ where $z_H=1/2\pi T$, and unit AdS radius. The fields $a$ and $\F$ are localised to the asymptotically $AdS_2$ subspace at $r=0$, with induced metric $g_{mn}$ ($m,n=z,t$). We describe the dynamics of $A$, $a$, and $\F$ by the simple quadratic action~\cite{Erdmenger:2013dpa},
\begin{subequations}
\label{action}
\begin{align}
	S&=-\frac{N}{4\pi}\int\limits_\textrm{BTZ}A\wedge\dif A+S_{AdS_2}\;,\\
	S_{AdS_2}&=-N\int\limits_{x=0}\dif z\dif t\sqrt{-g}\Big(\frac{1}{4}f^{mn}f_{mn}\\
	&\quad\quad\quad\quad\quad+\le(D^m\F\ri)^\dag\le(D_m\F\ri)+M^2\F^\dagger\F\Big),\nonumber
\end{align}
\end{subequations}
with field strength $f=\dif a$, covariant derivative $D_m\F=\le(\pa_m+iA_m-ia_m\ri)\F$, and mass-squared $M^2$. At the horizon $z=z_H$ we require regularity of all fields. At the boundary $z=0$, $a$'s leading mode, $a\sim Q/z$, is related to ${\cal Q}$: $Q\neq 0$ breaks $\chi$'s PHS, so the PHS value $Q=0$ is dual to the PHS value $\mathcal{Q}=N/2$, and increasing $|Q|$ corresponds to increasing $|\mathcal{Q}-N/2|$.

The large-$N$ Kondo interaction $-\lambda_K\mathcal{O}^\dagger\mathcal{O}/2$ is classically marginal, hence $\mathcal{O}$ has UV dimension $\Delta = 1/2$, which fixes $M^2$ and hence $\F$'s near-boundary expansion, $\F\sim\sqrt{z}\le(\alpha\log z+\beta\ri)$. Introducing the Kondo interaction amounts to adding a boundary term $\propto -\lambda_K\mathcal{O}^\dagger\mathcal{O}/2$ to $S$, which changes $\F$'s boundary condition from $\alpha=0$ to $\alpha=-\lambda_K\beta$~\cite{Witten:2001ua,Papadimitriou:2007sj,Erdmenger:2013dpa}. For more details about the boundary terms, see~\cite{O'Bannon:2015gwa,Erdmenger:2016jjg}. A holographic scaling analysis reveals that $\lambda_K$ runs logarithmically, $\lambda_K = 1/\log\left(T/T_K\right)$, diverging at the dynamically-generated Kondo temperature, $T_K \equiv \Lambda \, e^{-1/\lambda_K}/(2\pi)$, with $\lambda_K$ evaluated at the UV cutoff, $\Lambda$. A holographic antiferromagnetic UV Kondo coupling, $\lambda_K>0$, is thus marginally relevant, breaks conformal invariance, and triggers an RG flow.

As mentioned above, our model has a mean-field phase transition~\cite{Erdmenger:2013dpa}: $\langle \mathcal{O}\rangle=0$ ($\Phi=0$) when $T>T_c$ and $\langle \mathcal{O} \rangle \neq 0$ ($\Phi\neq0$) when $T \leq T_c$. Condensate formation $\langle \mathcal{O} \rangle \neq0$ breaks the charge and auxiliary $U(1)$'s to the diagonal, and signals the Kondo effect, including a phase shift of $\psi_L$, dual to a Wilson line of $A$, and impurity screening, dual to reduction of $f$ flux between $z=0$ and $z=z_H$. We refer to the $T>T_c$ and $T \leq T_c$ phases as ``unscreened'' and ``screened,'' respectively. In~\cite{Erdmenger:2013dpa,O'Bannon:2015gwa,Erdmenger:2015spo,Erdmenger:2015xpq} we computed $T_c$ numerically. Below we obtain an exact formula for $T_c$.

\vspace{-0.5cm}
\section{Fano Resonances}
\vspace{-0.3cm}
If a retarded Green's function of complex frequency $\omega$, $G(\w)$, has a pole at $\omega_p$, $G(\omega) \sim \frac{Z}{\omega-\omega_p}$, with \textit{complex} residue $Z = Z_R + i Z_I$, then near the pole the spectral function $\rho(\w) \equiv - 2\,\textrm{Im}(G(\w))$ will have a Fano resonance~\cite{PhysRev.124.1866,RevModPhys.82.2257} (setting $\textrm{Im}(\w)=0$),
\beq
\label{eq:fanospec}
\rho_{Fano}(\omega) = \frac{(\w - \w_0 + q\,\Gm/2)^2}{(\w - \w_0)^2 + (\Gm/2)^2} \,,  
\eeq
with position $\omega_0 = \textrm{Re}(\omega_p)$, width $\Gamma = 2|\textrm{Im}(\omega_p)|$, and \textit{asymmetry parameter} $q = -Z_R/Z_I+\sqrt{1+Z_R^2/Z_I^2}$. Fano resonances are anti-symmetric when $q=1$, meaning $\rho(\omega)$ is odd under PHS, and symmetric when $q=0$ (an anti-resonance) or $\infty$ (a Lorentzian), meaning $\rho(\omega)$ is even. Fano resonances arise when a Lorentzian resonance is immersed in a continuum (in energy), due to interference between the two. The asymmetry parameter $q$ contains key dynamical information, specifically, $q^2\propto$ the ratio of probabilities of resonant and non-resonant scattering.

In our model, the $AdS_2$ subspace inherits scale invariance from $AdS_3$, or in dual field theory language, the impurity inherits scale invariance from the CFT, so $\rho(\w)$ of impurity operators must be a featureless continuum. Our marginally-relevant Kondo coupling then breaks scale invariance and produces a resonance, while $Q\neq 0$ breaks PHS. We will show that $\rho(\w)$ of $\mathcal{O}$ then indeed generically exhibits asymmetric Fano resonances.

\vspace{-0.5cm}
\section{Spectral Functions}
\vspace{-0.3cm}
We compute $G(\w)$ holographically by solving for linearized fluctuations about solutions for the unscreened and screened phases~\cite{Son:2002sd,Kovtun:2005ev,Erdmenger:2016jjg}. At all $T$, we find that the Kac-Moody current's $G(\w)$ is unaffected by the impurity. In the unscreened phase, we find that all charged $G(\w)$ vanish, \textit{i.e.}\ $\goo(\w) \equiv \langle \mathcal{O}(\omega) \mathcal{O}(-\omega)\rangle=0$, while 
\beq
G_{\mathcal{O}^\hc\mathcal{O}}(\w) \equiv \langle \mathcal{O}^{\dagger}(\omega) \mathcal{O}(-\omega)\rangle = \frac{N}{\lambda_K}\left(1 - \frac{1}{\lambda_K D(\w)}\right), \nonumber
\eeq
\vspace{-0.5cm}
\beq
D(\omega) \equiv H\left[-\frac{1}{2} + i Q - \frac{i \omega}{2 \pi T}\right] + H\left[-\frac{1}{2} - i Q\right] + \ln\left(\frac{2T}{T_K}\right), \nonumber
\eeq
with Harmonic number $H[x]$, and $\lambda_K$ evaluated at $\Lambda$. The form of $\good(\w)$ is the same, but with $Q \to -Q$. Scale invariance in $(0+1)$-dimensions and $\Delta =1/2$ imply a trivial UV continuum: $\lim_{\w\to \infty}\rodo(\w) = 0$.

\begin{figure}[htpb]
\centering
\includegraphics[width=3.3in]{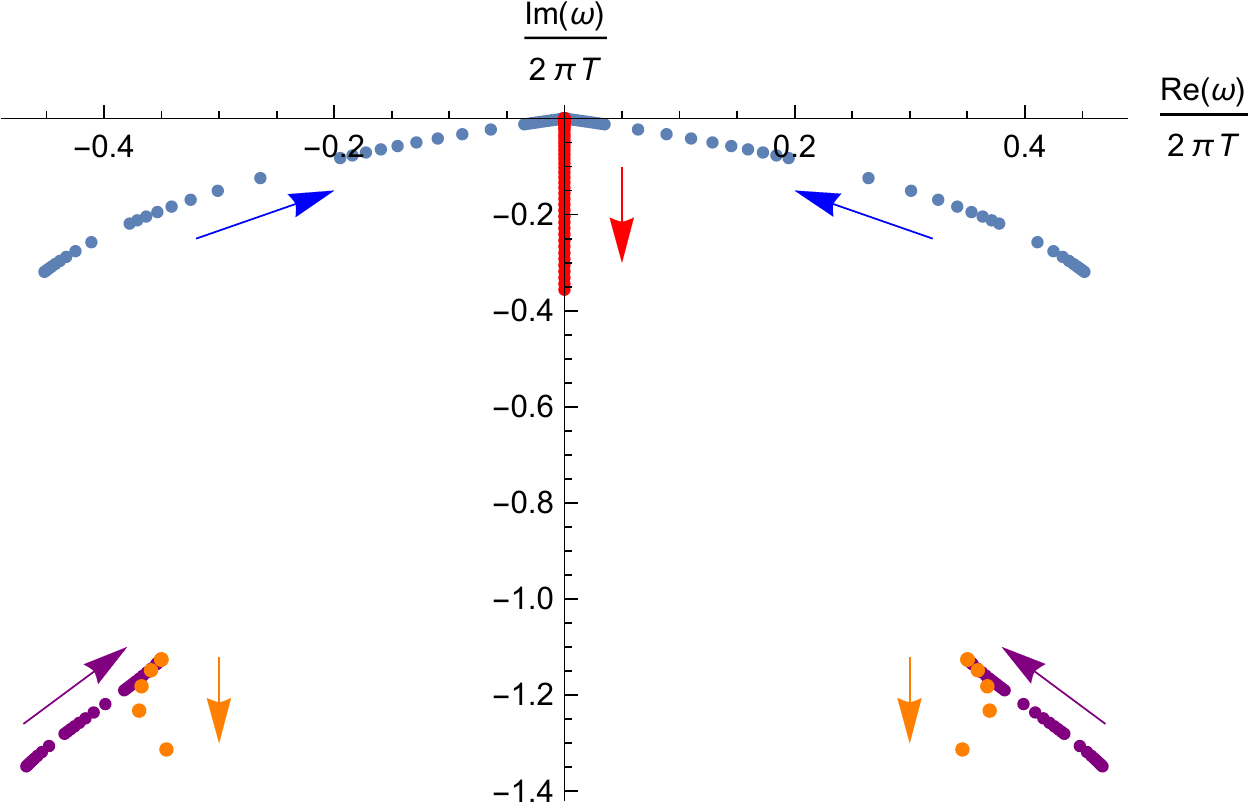} 
\caption{Positions of poles in the complex $\w/(2\pi T)$ plane for $Q = 1/2$. Blue and purple denote lowest and next-lowest poles, respectively, of $\godo(\w)$ ($\textrm{Re}(\w)>0$) and $\good(\w)$ ($\textrm{Re}(\w)<0$), for $T/T_c$ from $100$ down to $1.001$. Red and orange denote the same for $\godo(\w)$ for $T/T_c$ from $1$ down to $0.2$ Arrows indicate movement of poles as $T$ decreases.}
\label{fig:QNMs}
\end{figure}

For given $Q$ and $T$, $\godo(\omega)$ has poles in $\w$ when $D(\omega)=0$. Fig.~\ref{fig:QNMs} shows our numerical results for the positions of the lowest (closest to $\w=0$) and next-lowest poles of $\godo(\w)$ and $\good(\w)$ in the complex $\omega/(2 \pi T)$ plane, for $Q=1/2$. Other $Q$ give similar results. As $T \to T_c^+$, the lowest pole moves towards the origin, arrives there at $T_c$, and when $T<T_c$, crosses into the $\textrm{Im}(\omega)>0$ region, signaling instability (not shown). We thus identify $T_c$ as the $T$ where $D(\omega=0)=0$,
\beq
T_c = \frac{1}{2} \, T_K \, \textrm{exp} \left [-2\,\mathrm{Re}\left(H\left[\frac{1}{2} + i Q\right] \right)\right]. \nonumber
\eeq

\begin{figure}[htpb]
\centering
\includegraphics[width=3.3in]{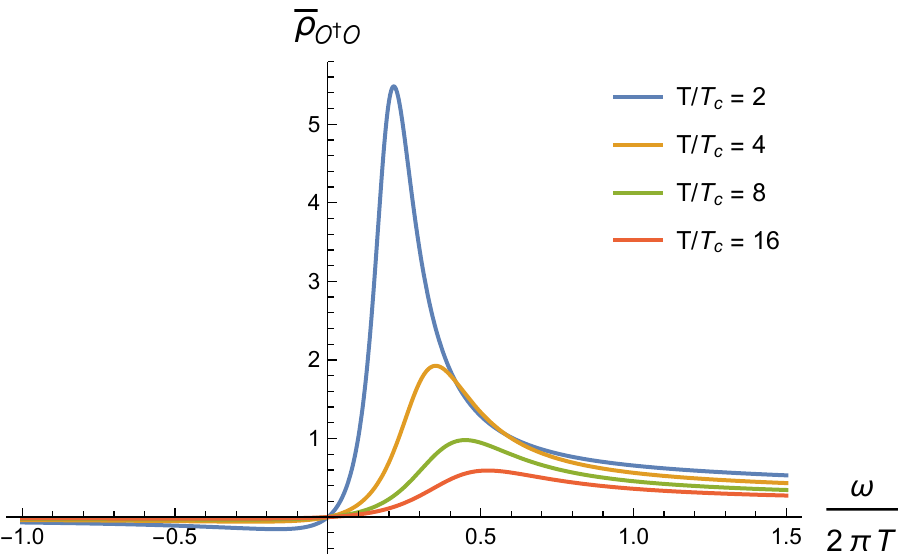}  
\caption{The normalized spectral function $\bar{\rho}_{\mcO^\hc\mcO}(\w)$ versus real $\w/(2\pi T)$ for $Q = 1/2$ and, from shortest to tallest, $T/T_c = 16$ (red), $8$ (green), $4$ (orange), and $2$ (blue).}
\label{fig:ASFano}
\end{figure}

Fig.~\ref{fig:ASFano} shows the normalized spectral function $\bar{\rho}_{\mathcal{O}^\hc\mathcal{O}}(\w) \equiv -2\frac{\lambda_K^2}{N}\,\mathrm{Im}\,G_{\mathcal{O}^\hc\mathcal{O}}(\w)$ versus real $\omega/(2 \pi T)$ for $Q=1/2$ and $T/T_c=16,8,4,2$. We find a Fano resonance, as advertised, with asymmetric minimum and maximum. Numerically, $\omega_0\approx \textrm{Re}(\omega_p)$ and $\Gamma\approx 2|\textrm{Im}(\omega_p)|$, as in~\eqref{eq:fanospec}, where $\w_p$ is $\godo(\w)$'s lowest pole. As $T \to T_c^+$, $q$ grows: $q \approx 1.7$ at $T=16T_c$ while $q \approx 4$ at $T= 2T_c$.

For $T$ just above $T_c$, $T \gtrsim T_c$, expanding in $T$ about $T_c$ and in $\w$ about $\w=0$ gives, for $\godo(\w)$'s lowest pole,
\beq
\label{eq:lowestQNM}
\frac{\w_p}{2\pi T} = -i\frac{T/T_c - 1}{\psi'[\half + i Q]} \,, \quad
Z = -i\frac{N}{\lambda_K^2}\frac{2\pi T_c}{\psi'[\half + i Q]},
\eeq
with digamma function $\psi[x]$. The resonance height thus grows as $(T/T_c-1)^{-1}$ and the width shrinks as $T/T_c-1$. It is therefore \textit{not} related to a Kondo resonance, which grows \textit{logarithmically} as $T \to T_K^+$~\cite{Phillips2012}. Indeed, at large $N$ we expect the Kondo resonance only in the screened phase~\cite{Coleman2015}. Our resonance is presumably a bound state of $\psi_L$ and $\chi$, heralding the nascent screened phase.

\begin{figure}[htpb]
\centering
\includegraphics[width=3in]{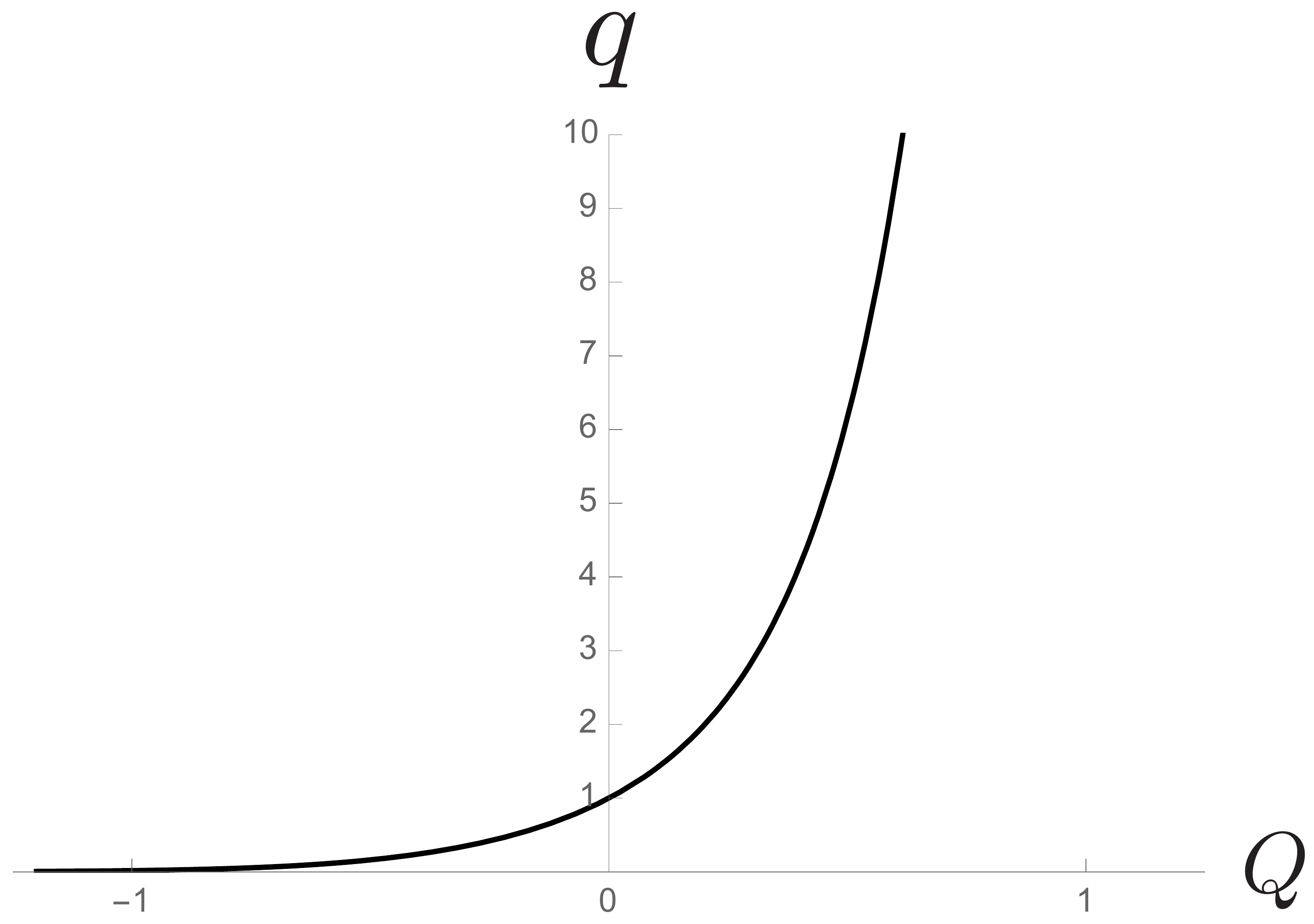} 
\caption{Asymmetry parameter $q$ versus $Q$, for $T\gtrsim T_c$.}
\label{fig:qvsQ}
\end{figure}

\begin{figure}[htpb]
\centering
\includegraphics[width=3.3in]{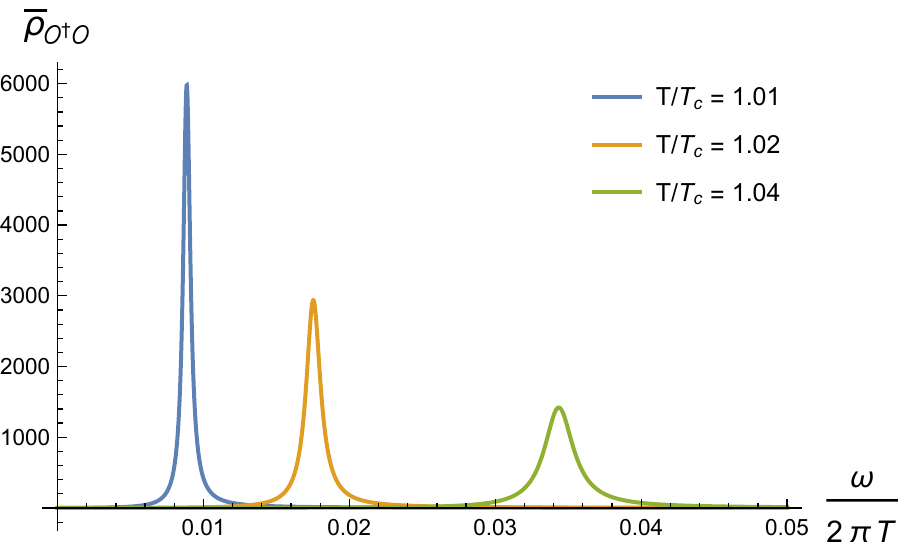} 
\caption{The normalized spectral function, $\bar{\rho}_{\mcO^\hc\mcO}(\w)$, versus real $\w/(2\pi T)$ for $Q = 1$ and, from shortest to tallest, $T/T_c=1.04$ (green), $1.02$ (orange), and $1.01$ (blue).}
\label{fig:SFano}
\end{figure}

The $Z$ in~\eqref{eq:lowestQNM} gives $q$ that depends only on $Q$, shown in fig.~\ref{fig:qvsQ}. (Anti-)symmetric values $q=1,0, \infty$ occur when $Q \to 0,\mp \infty$, respectively. Indeed, fig.~\ref{fig:SFano} shows that even for relatively modest $Q=1$, the resonance is nearly Lorentzian, the minimum having practically vanished.

In the screened phase, we have numerical results for $\godo(\w)$~\cite{Erdmenger:2013dpa,O'Bannon:2015gwa,Erdmenger:2015spo,Erdmenger:2015xpq,Erdmenger:2016jjg}. Fig.~\ref{fig:QNMs} shows our numerical results for the positions of the lowest and next-lowest poles in $\godo(\w)$ for $Q=1/2$. Other $Q$ give similar results. At $T=T_c$ the poles are coincident with those of $\godo(\w)$ and $\good(\w)$ in the unscreened phase. As $T$ decreases below $T_c$, $\godo(\w)$'s lowest pole, $\w_p$, moves straight down the $\textrm{Im}(\w)$ axis.

\begin{figure}[htpb]
\centering
\includegraphics[width=3.4in]{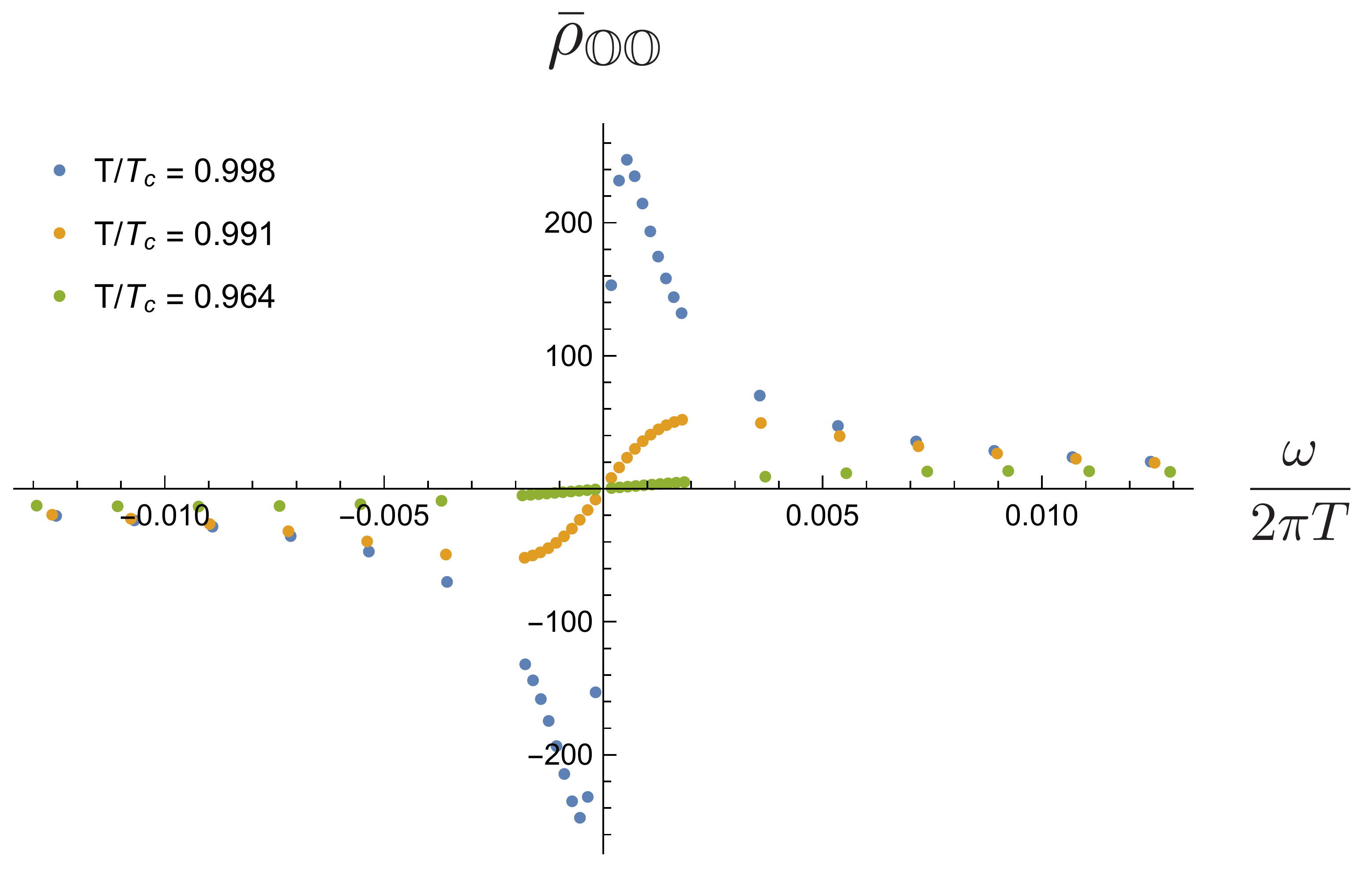} 
\caption{The normalized spectral function $\bar{\rho}_{\mathcal{O}^{\dagger}\mathcal{O}}(\w)$ versus real $\w/(2\pi T)$ for $Q = 1/2$ and, from tallest to shortest, $T/T_c =0.998$ (blue), $0.991$ (orange), and $0.964$ (green).}
\label{fig:SFS}
\end{figure}

From our experience with the unscreened phase, we expect $\w_p$ to produce a Fano resonance in the normalized spectral function, $\bar{\rho}_{\mathcal{O}^{\dagger}\mathcal{O}}(\w)$. Crucially, $\textrm{Re}(w_p)=0$, so $\w_p$ preserves PHS, $\textrm{Re}(\w) \to - \textrm{Re}(\w)$, so we expect an \textit{anti-symmetric} Fano resonance at $\textrm{Re}(\w)=0$. Moreover, $|\textrm{Im}(\w_p)|$ increases as $T$ decreases, and so should the width $\Gamma$. Fig.~\ref{fig:SFS} confirms our expectations: $\bar{\rho}_{\mathcal{O}^{\dagger}\mathcal{O}}(\w)$'s only significant feature is a Fano resonance at $\textrm{Re}(\w)=0$ with $q=1$, meaning perfectly anti-symmetric minimum and maximum, and whose $\Gamma$ increases as $T$ decreases. Additionally, the height decreases, and indeed our numerics suggest $\lim_{T\to0}\bar{\rho}_{\mathcal{O}^{\dagger}\mathcal{O}}(\w)=0$.

\begin{figure}[htpb]
\centering
\includegraphics[width=3.4in]{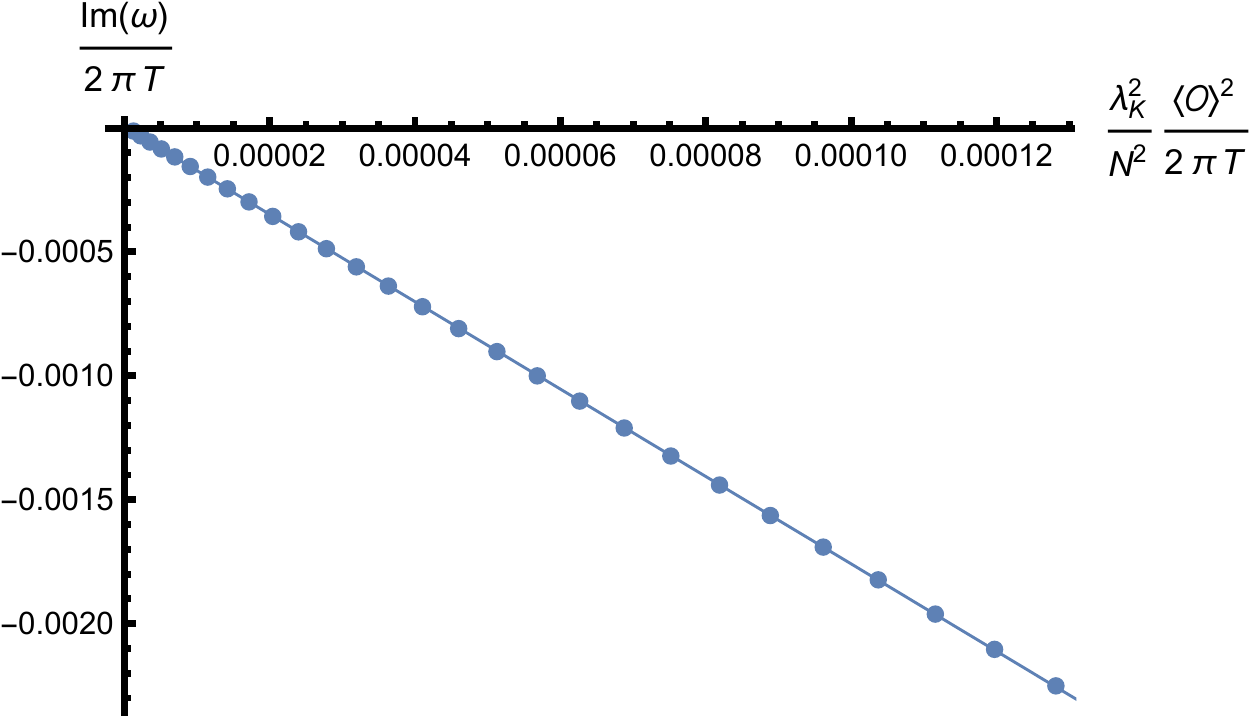} 
\caption{Dots denote the position of $\godo(\w)$'s lowest pole versus $\frac{\lambda_K^2}{N^2}\langle\mcO\rangle^2/(2 \pi T)$ for $Q = 1/2$. The solid line is a linear fit with slope $\approx - 17.6$ and intercept at the origin.}
\label{fig:wOvev}
\end{figure}

Fig.~\ref{fig:wOvev} shows our numerical results for the position of $\w_p$ versus small $\frac{\lambda_K^2}{N^2}\langle\mcO\rangle^2/(2 \pi T)$, or equivalently, $T$ just below $T_c$, $T \lesssim T_c$, for $Q = 1/2$. Fig.~\ref{fig:wOvev} also shows a linear fit demonstrating that\footnote{In~\cite{Erdmenger:2016jjg} we derive~\eqref{eq:kondopole} without numerics, via a small-$\langle \mathcal{O}\rangle$ expansion.}
\beq
\w_p \propto -i \left\langle {\cal O} \right\rangle^2.
\label{eq:kondopole}
\eeq
Our model's mean-field behavior $\langle \mathcal{O}\rangle \propto (T_c-T)^{1/2}$ then implies $\Gamma\propto T_c-T$ for $T \lesssim T_c$. 

The behavior in~\eqref{eq:kondopole} is in fact identical to that in a LFL at large $N$. In a LFL, the Kondo resonance is formally defined in the LFL fermion spectral function, and at large $N$ appears only in the screened phase, with $\Gamma \propto \langle\mathcal{O}\rangle^2$~\cite{Coleman2015}. For $T \lesssim T_c$, the mean-field behavior $\langle \mathcal{O}\rangle\propto (T_c-T)^{1/2}$ then implies $\Gamma\propto T_c-T$. Crucially, in the screened phase the Kondo resonance also appears in other spectral functions, due to operator mixing induced by the symmetry breaking~\cite{Coleman2015}. In particular, a Kondo resonance should produce a pole in $\godo(\w)$ precisely of the form in~\eqref{eq:kondopole}\footnote{For details, see for example chapter 18 of~\cite{Coleman2015}.}. Our result~\eqref{eq:kondopole} thus proves the existence of a Kondo resonance in our model when $T \lesssim T_c$, with defining features essentially intact despite the strong interactions.

\vspace{-0.5cm}
\section{Conclusion}
\vspace{-0.3cm}

In a holographic model describing the interaction of a magnetic impurity with a strongly correlated CFT at large $N$, we discovered a novel mechanism for producing Fano resonances, namely via RG flows between $(0+1)$-dimensional fixed points. The origin and consequences of such Fano resonances, in existing cases that have gone unidentified and in novel cases, deserve further study, particularly of the physics contained in the asymmetry parameter $q$.

\section*{Acknowledgments}

We would like to thank Ian Affleck, Natan Andrei, Piers Coleman, Mario Flory, Henrik Johannesson, Andrew Mitchell, Max Newrzella, and Philip Phillips for helpful conversations and correspondence. C.H. is supported by the Ramon y Cajal fellowship RYC-2011-07593, the Asturian grant FC-15-GRUPIN14-108 and the Spanish national grant MINECO-16-FPA2015-63667-P. A.~O'B. is a Royal Society University Research Fellow. J.~P. is supported by the Clarendon Fund and St John's College, Oxford, and by the European Research Council under the European Union's Seventh Framework Programme (ERC Grant agreement 307955).

\bibliography{fanoletterv3}

\begin{thebibliography}{52}%
\makeatletter
\providecommand \@ifxundefined [1]{%
 \@ifx{#1\undefined}
}%
\providecommand \@ifnum [1]{%
 \ifnum #1\expandafter \@firstoftwo
 \else \expandafter \@secondoftwo
 \fi
}%
\providecommand \@ifx [1]{%
 \ifx #1\expandafter \@firstoftwo
 \else \expandafter \@secondoftwo
 \fi
}%
\providecommand \natexlab [1]{#1}%
\providecommand \enquote  [1]{``#1''}%
\providecommand \bibnamefont  [1]{#1}%
\providecommand \bibfnamefont [1]{#1}%
\providecommand \citenamefont [1]{#1}%
\providecommand \href@noop [0]{\@secondoftwo}%
\providecommand \href [0]{\begingroup \@sanitize@url \@href}%
\providecommand \@href[1]{\@@startlink{#1}\@@href}%
\providecommand \@@href[1]{\endgroup#1\@@endlink}%
\providecommand \@sanitize@url [0]{\catcode `\\12\catcode `\$12\catcode
  `\&12\catcode `\#12\catcode `\^12\catcode `\_12\catcode `\%12\relax}%
\providecommand \@@startlink[1]{}%
\providecommand \@@endlink[0]{}%
\providecommand \url  [0]{\begingroup\@sanitize@url \@url }%
\providecommand \@url [1]{\endgroup\@href {#1}{\urlprefix }}%
\providecommand \urlprefix  [0]{URL }%
\providecommand \Eprint [0]{\href }%
\providecommand \doibase [0]{http://dx.doi.org/}%
\providecommand \selectlanguage [0]{\@gobble}%
\providecommand \bibinfo  [0]{\@secondoftwo}%
\providecommand \bibfield  [0]{\@secondoftwo}%
\providecommand \translation [1]{[#1]}%
\providecommand \BibitemOpen [0]{}%
\providecommand \bibitemStop [0]{}%
\providecommand \bibitemNoStop [0]{.\EOS\space}%
\providecommand \EOS [0]{\spacefactor3000\relax}%
\providecommand \BibitemShut  [1]{\csname bibitem#1\endcsname}%
\let\auto@bib@innerbib\@empty
\bibitem [{\citenamefont {Kondo}(1964)}]{PTP.32.37}%
  \BibitemOpen
  \bibfield  {author} {\bibinfo {author} {\bibfnamefont {J.}~\bibnamefont
  {Kondo}},\ }\href {\doibase 10.1143/PTP.32.37} {\bibfield  {journal}
  {\bibinfo  {journal} {Prog. Theo. Phys.}\ }\textbf {\bibinfo {volume} {32}},\
  \bibinfo {pages} {37} (\bibinfo {year} {1964})}\BibitemShut {NoStop}%
\bibitem [{\citenamefont {Hewson}(1993)}]{Hewson:1993}%
  \BibitemOpen
  \bibfield  {author} {\bibinfo {author} {\bibfnamefont {A.}~\bibnamefont
  {Hewson}},\ }\href@noop {} {\bibfield  {journal} {\bibinfo  {journal}
  {{Cambridge University Press}}\ } (\bibinfo {year} {1993})}\BibitemShut
  {NoStop}%
\bibitem [{\citenamefont {Cox}\ and\ \citenamefont
  {Zawadowski}(1998)}]{doi:10.1080/000187398243500}%
  \BibitemOpen
  \bibfield  {author} {\bibinfo {author} {\bibfnamefont {D.~L.}\ \bibnamefont
  {Cox}}\ and\ \bibinfo {author} {\bibfnamefont {A.}~\bibnamefont
  {Zawadowski}},\ }\href {\doibase 10.1080/000187398243500} {\bibfield
  {journal} {\bibinfo  {journal} {Advances in Physics}\ }\textbf {\bibinfo
  {volume} {47}},\ \bibinfo {pages} {599} (\bibinfo {year} {1998})},\ \Eprint
  {http://arxiv.org/abs/{arxiv:cond-mat/9704103}} {{arxiv:cond-mat/9704103}}
  \BibitemShut {NoStop}%
\bibitem [{\citenamefont {Lee}\ and\ \citenamefont
  {Toner}(1992)}]{PhysRevLett.69.3378}%
  \BibitemOpen
  \bibfield  {author} {\bibinfo {author} {\bibfnamefont {D.-H.}\ \bibnamefont
  {Lee}}\ and\ \bibinfo {author} {\bibfnamefont {J.}~\bibnamefont {Toner}},\
  }\href {\doibase 10.1103/PhysRevLett.69.3378} {\bibfield  {journal} {\bibinfo
   {journal} {Phys. Rev. Lett.}\ }\textbf {\bibinfo {volume} {69}},\ \bibinfo
  {pages} {3378} (\bibinfo {year} {1992})}\BibitemShut {NoStop}%
\bibitem [{\citenamefont {Furusaki}\ and\ \citenamefont
  {Nagaosa}(1994)}]{PhysRevLett.72.892}%
  \BibitemOpen
  \bibfield  {author} {\bibinfo {author} {\bibfnamefont {A.}~\bibnamefont
  {Furusaki}}\ and\ \bibinfo {author} {\bibfnamefont {N.}~\bibnamefont
  {Nagaosa}},\ }\href {\doibase 10.1103/PhysRevLett.72.892} {\bibfield
  {journal} {\bibinfo  {journal} {Phys. Rev. Lett.}\ }\textbf {\bibinfo
  {volume} {72}},\ \bibinfo {pages} {892} (\bibinfo {year} {1994})}\BibitemShut
  {NoStop}%
\bibitem [{\citenamefont {Fr\"ojdh}\ and\ \citenamefont
  {Johannesson}(1995)}]{PhysRevLett.75.300}%
  \BibitemOpen
  \bibfield  {author} {\bibinfo {author} {\bibfnamefont {P.}~\bibnamefont
  {Fr\"ojdh}}\ and\ \bibinfo {author} {\bibfnamefont {H.}~\bibnamefont
  {Johannesson}},\ }\href {\doibase 10.1103/PhysRevLett.75.300} {\bibfield
  {journal} {\bibinfo  {journal} {Phys. Rev. Lett.}\ }\textbf {\bibinfo
  {volume} {75}},\ \bibinfo {pages} {300} (\bibinfo {year} {1995})}\BibitemShut
  {NoStop}%
\bibitem [{\citenamefont {Fr\"ojdh}\ and\ \citenamefont
  {Johannesson}(1996)}]{PhysRevB.53.3211}%
  \BibitemOpen
  \bibfield  {author} {\bibinfo {author} {\bibfnamefont {P.}~\bibnamefont
  {Fr\"ojdh}}\ and\ \bibinfo {author} {\bibfnamefont {H.}~\bibnamefont
  {Johannesson}},\ }\href {\doibase 10.1103/PhysRevB.53.3211} {\bibfield
  {journal} {\bibinfo  {journal} {Phys. Rev. B}\ }\textbf {\bibinfo {volume}
  {53}},\ \bibinfo {pages} {3211} (\bibinfo {year} {1996})}\BibitemShut
  {NoStop}%
\bibitem [{\citenamefont {{Furusaki}}(2005)}]{2005JPSJ...74...73F}%
  \BibitemOpen
  \bibfield  {author} {\bibinfo {author} {\bibfnamefont {A.}~\bibnamefont
  {{Furusaki}}},\ }\href {\doibase 10.1143/JPSJ.74.73} {\bibfield  {journal}
  {\bibinfo  {journal} {Journal of the Physical Society of Japan}\ }\textbf
  {\bibinfo {volume} {74}},\ \bibinfo {pages} {73} (\bibinfo {year} {2005})},\
  \Eprint {http://arxiv.org/abs/cond-mat/0409016} {cond-mat/0409016}
  \BibitemShut {NoStop}%
\bibitem [{\citenamefont {Fulde}\ \emph {et~al.}(1993)\citenamefont {Fulde},
  \citenamefont {Zevin},\ and\ \citenamefont {Zwicknagl}}]{Fulde1993}%
  \BibitemOpen
  \bibfield  {author} {\bibinfo {author} {\bibfnamefont {P.}~\bibnamefont
  {Fulde}}, \bibinfo {author} {\bibfnamefont {V.}~\bibnamefont {Zevin}}, \ and\
  \bibinfo {author} {\bibfnamefont {G.}~\bibnamefont {Zwicknagl}},\ }\href
  {\doibase 10.1007/BF01312167} {\bibfield  {journal} {\bibinfo  {journal}
  {Zeitschrift f{\"u}r Physik B Condensed Matter}\ }\textbf {\bibinfo {volume}
  {92}},\ \bibinfo {pages} {133} (\bibinfo {year} {1993})}\BibitemShut
  {NoStop}%
\bibitem [{\citenamefont {{Schork}}\ and\ \citenamefont
  {{Fulde}}(1994)}]{1994PhRvB..50.1345S}%
  \BibitemOpen
  \bibfield  {author} {\bibinfo {author} {\bibfnamefont {T.}~\bibnamefont
  {{Schork}}}\ and\ \bibinfo {author} {\bibfnamefont {P.}~\bibnamefont
  {{Fulde}}},\ }\href {\doibase 10.1103/PhysRevB.50.1345} {\bibfield  {journal}
  {\bibinfo  {journal} {\prb}\ }\textbf {\bibinfo {volume} {50}},\ \bibinfo
  {pages} {1345} (\bibinfo {year} {1994})},\ \Eprint
  {http://arxiv.org/abs/cond-mat/9402106} {cond-mat/9402106} \BibitemShut
  {NoStop}%
\bibitem [{\citenamefont {Khaliullin}\ and\ \citenamefont
  {Fulde}(1995)}]{PhysRevB.52.9514}%
  \BibitemOpen
  \bibfield  {author} {\bibinfo {author} {\bibfnamefont {G.}~\bibnamefont
  {Khaliullin}}\ and\ \bibinfo {author} {\bibfnamefont {P.}~\bibnamefont
  {Fulde}},\ }\href {\doibase 10.1103/PhysRevB.52.9514} {\bibfield  {journal}
  {\bibinfo  {journal} {Phys. Rev. B}\ }\textbf {\bibinfo {volume} {52}},\
  \bibinfo {pages} {9514} (\bibinfo {year} {1995})}\BibitemShut {NoStop}%
\bibitem [{\citenamefont {Igarashi}\ \emph {et~al.}(1995)\citenamefont
  {Igarashi}, \citenamefont {Murayama},\ and\ \citenamefont
  {Fulde}}]{PhysRevB.52.15966}%
  \BibitemOpen
  \bibfield  {author} {\bibinfo {author} {\bibfnamefont {J.}~\bibnamefont
  {Igarashi}}, \bibinfo {author} {\bibfnamefont {K.}~\bibnamefont {Murayama}},
  \ and\ \bibinfo {author} {\bibfnamefont {P.}~\bibnamefont {Fulde}},\ }\href
  {\doibase 10.1103/PhysRevB.52.15966} {\bibfield  {journal} {\bibinfo
  {journal} {Phys. Rev. B}\ }\textbf {\bibinfo {volume} {52}},\ \bibinfo
  {pages} {15966} (\bibinfo {year} {1995})}\BibitemShut {NoStop}%
\bibitem [{\citenamefont {{Tornow}}\ \emph {et~al.}(1997)\citenamefont
  {{Tornow}}, \citenamefont {{Zevin}},\ and\ \citenamefont
  {{Zwicknagl}}}]{1997cond.mat..1137T}%
  \BibitemOpen
  \bibfield  {author} {\bibinfo {author} {\bibfnamefont {S.}~\bibnamefont
  {{Tornow}}}, \bibinfo {author} {\bibfnamefont {V.}~\bibnamefont {{Zevin}}}, \
  and\ \bibinfo {author} {\bibfnamefont {G.}~\bibnamefont {{Zwicknagl}}},\
  }\href@noop {} {\bibfield  {journal} {\bibinfo  {journal} {eprint
  arXiv:cond-mat/9701137}\ } (\bibinfo {year} {1997})},\ \Eprint
  {http://arxiv.org/abs/cond-mat/9701137} {cond-mat/9701137} \BibitemShut
  {NoStop}%
\bibitem [{\citenamefont {{Schork}}\ and\ \citenamefont
  {{Blawid}}(1997)}]{1997PhRvB..56.6559S}%
  \BibitemOpen
  \bibfield  {author} {\bibinfo {author} {\bibfnamefont {T.}~\bibnamefont
  {{Schork}}}\ and\ \bibinfo {author} {\bibfnamefont {S.}~\bibnamefont
  {{Blawid}}},\ }\href {\doibase 10.1103/PhysRevB.56.6559} {\bibfield
  {journal} {\bibinfo  {journal} {\prb}\ }\textbf {\bibinfo {volume} {56}},\
  \bibinfo {pages} {6559} (\bibinfo {year} {1997})},\ \Eprint
  {http://arxiv.org/abs/cond-mat/9706225} {cond-mat/9706225} \BibitemShut
  {NoStop}%
\bibitem [{\citenamefont {{Davidovich}}\ and\ \citenamefont
  {{Zevin}}(1998)}]{1998PhRvB..57.7773D}%
  \BibitemOpen
  \bibfield  {author} {\bibinfo {author} {\bibfnamefont {B.}~\bibnamefont
  {{Davidovich}}}\ and\ \bibinfo {author} {\bibfnamefont {V.}~\bibnamefont
  {{Zevin}}},\ }\href {\doibase 10.1103/PhysRevB.57.7773} {\bibfield  {journal}
  {\bibinfo  {journal} {\prb}\ }\textbf {\bibinfo {volume} {57}},\ \bibinfo
  {pages} {7773} (\bibinfo {year} {1998})},\ \Eprint
  {http://arxiv.org/abs/cond-mat/9706283} {cond-mat/9706283} \BibitemShut
  {NoStop}%
\bibitem [{\citenamefont {Duc}\ and\ \citenamefont
  {Thang}(1999)}]{doi:10.1142/S0217984999001044}%
  \BibitemOpen
  \bibfield  {author} {\bibinfo {author} {\bibfnamefont {H.~T.}\ \bibnamefont
  {Duc}}\ and\ \bibinfo {author} {\bibfnamefont {N.~T.}\ \bibnamefont
  {Thang}},\ }\href {\doibase 10.1142/S0217984999001044} {\bibfield  {journal}
  {\bibinfo  {journal} {Modern Physics Letters B}\ }\textbf {\bibinfo {volume}
  {13}},\ \bibinfo {pages} {849} (\bibinfo {year} {1999})}\BibitemShut
  {NoStop}%
\bibitem [{\citenamefont {{Hofstetter}}\ \emph {et~al.}(2000)\citenamefont
  {{Hofstetter}}, \citenamefont {{Bulla}},\ and\ \citenamefont
  {{Vollhardt}}}]{2000PhRvL..84.4417H}%
  \BibitemOpen
  \bibfield  {author} {\bibinfo {author} {\bibfnamefont {W.}~\bibnamefont
  {{Hofstetter}}}, \bibinfo {author} {\bibfnamefont {R.}~\bibnamefont
  {{Bulla}}}, \ and\ \bibinfo {author} {\bibfnamefont {D.}~\bibnamefont
  {{Vollhardt}}},\ }\href {\doibase 10.1103/PhysRevLett.84.4417} {\bibfield
  {journal} {\bibinfo  {journal} {Physical Review Letters}\ }\textbf {\bibinfo
  {volume} {84}},\ \bibinfo {pages} {4417} (\bibinfo {year} {2000})},\ \Eprint
  {http://arxiv.org/abs/cond-mat/9912396} {cond-mat/9912396} \BibitemShut
  {NoStop}%
\bibitem [{\citenamefont {Erdmenger}\ \emph {et~al.}(2013)\citenamefont
  {Erdmenger}, \citenamefont {Hoyos}, \citenamefont {O'Bannon},\ and\
  \citenamefont {Wu}}]{Erdmenger:2013dpa}%
  \BibitemOpen
  \bibfield  {author} {\bibinfo {author} {\bibfnamefont {J.}~\bibnamefont
  {Erdmenger}}, \bibinfo {author} {\bibfnamefont {C.}~\bibnamefont {Hoyos}},
  \bibinfo {author} {\bibfnamefont {A.}~\bibnamefont {O'Bannon}}, \ and\
  \bibinfo {author} {\bibfnamefont {J.}~\bibnamefont {Wu}},\ }\href {\doibase
  10.1007/JHEP12(2013)086} {\bibfield  {journal} {\bibinfo  {journal} {JHEP}\
  }\textbf {\bibinfo {volume} {12}},\ \bibinfo {pages} {086} (\bibinfo {year}
  {2013})},\ \Eprint {http://arxiv.org/abs/1310.3271} {arXiv:1310.3271
  [hep-th]} \BibitemShut {NoStop}%
\bibitem [{\citenamefont {O'Bannon}\ \emph {et~al.}(2016)\citenamefont
  {O'Bannon}, \citenamefont {Papadimitriou},\ and\ \citenamefont
  {Probst}}]{O'Bannon:2015gwa}%
  \BibitemOpen
  \bibfield  {author} {\bibinfo {author} {\bibfnamefont {A.}~\bibnamefont
  {O'Bannon}}, \bibinfo {author} {\bibfnamefont {I.}~\bibnamefont
  {Papadimitriou}}, \ and\ \bibinfo {author} {\bibfnamefont {J.}~\bibnamefont
  {Probst}},\ }\href {\doibase 10.1007/JHEP01(2016)103} {\bibfield  {journal}
  {\bibinfo  {journal} {JHEP}\ }\textbf {\bibinfo {volume} {01}},\ \bibinfo
  {pages} {103} (\bibinfo {year} {2016})},\ \Eprint
  {http://arxiv.org/abs/1510.08123} {arXiv:1510.08123 [hep-th]} \BibitemShut
  {NoStop}%
\bibitem [{\citenamefont {Erdmenger}\ \emph {et~al.}(2016)\citenamefont
  {Erdmenger}, \citenamefont {Flory}, \citenamefont {Hoyos}, \citenamefont
  {Newrzella},\ and\ \citenamefont {Wu}}]{Erdmenger:2015spo}%
  \BibitemOpen
  \bibfield  {author} {\bibinfo {author} {\bibfnamefont {J.}~\bibnamefont
  {Erdmenger}}, \bibinfo {author} {\bibfnamefont {M.}~\bibnamefont {Flory}},
  \bibinfo {author} {\bibfnamefont {C.}~\bibnamefont {Hoyos}}, \bibinfo
  {author} {\bibfnamefont {M.-N.}\ \bibnamefont {Newrzella}}, \ and\ \bibinfo
  {author} {\bibfnamefont {J.~M.~S.}\ \bibnamefont {Wu}},\ }\href {\doibase
  10.1002/prop.201500099} {\bibfield  {journal} {\bibinfo  {journal} {Fortsch.
  Phys.}\ }\textbf {\bibinfo {volume} {64}},\ \bibinfo {pages} {109} (\bibinfo
  {year} {2016})},\ \Eprint {http://arxiv.org/abs/1511.03666} {arXiv:1511.03666
  [hep-th]} \BibitemShut {NoStop}%
\bibitem [{\citenamefont {Erdmenger}\ \emph {et~al.}(2015)\citenamefont
  {Erdmenger}, \citenamefont {Flory}, \citenamefont {Hoyos}, \citenamefont
  {Newrzella}, \citenamefont {O'Bannon},\ and\ \citenamefont
  {Wu}}]{Erdmenger:2015xpq}%
  \BibitemOpen
  \bibfield  {author} {\bibinfo {author} {\bibfnamefont {J.}~\bibnamefont
  {Erdmenger}}, \bibinfo {author} {\bibfnamefont {M.}~\bibnamefont {Flory}},
  \bibinfo {author} {\bibfnamefont {C.}~\bibnamefont {Hoyos}}, \bibinfo
  {author} {\bibfnamefont {M.-N.}\ \bibnamefont {Newrzella}}, \bibinfo {author}
  {\bibfnamefont {A.}~\bibnamefont {O'Bannon}}, \ and\ \bibinfo {author}
  {\bibfnamefont {J.}~\bibnamefont {Wu}},\ }in\ \href
  {https://inspirehep.net/record/1407205/files/arXiv:1511.09362.pdf} {\emph
  {\bibinfo {booktitle} {{The String Theory Universe, 21st European String
  Workshop and 3rd COST MP1210 Meeting Leuven, Belgium, September 7-11,
  2015}}}}\ (\bibinfo {year} {2015})\ \Eprint {http://arxiv.org/abs/1511.09362}
  {arXiv:1511.09362 [hep-th]} \BibitemShut {NoStop}%
\bibitem [{\citenamefont {Phillips}(2012)}]{Phillips2012}%
  \BibitemOpen
  \bibfield  {author} {\bibinfo {author} {\bibfnamefont {P.}~\bibnamefont
  {Phillips}},\ }\href@noop {} {\bibfield  {journal} {\bibinfo  {journal}
  {{Cambridge University Press}}\ } (\bibinfo {year} {2012})}\BibitemShut
  {NoStop}%
\bibitem [{\citenamefont {Coleman}(2015)}]{Coleman2015}%
  \BibitemOpen
  \bibfield  {author} {\bibinfo {author} {\bibfnamefont {P.}~\bibnamefont
  {Coleman}},\ }\href@noop {} {\bibfield  {journal} {\bibinfo  {journal}
  {{Cambridge University Press}}\ } (\bibinfo {year} {2015})}\BibitemShut
  {NoStop}%
\bibitem [{\citenamefont {Coleman}\ and\ \citenamefont
  {Andrei}(1986)}]{0022-3719-19-17-017}%
  \BibitemOpen
  \bibfield  {author} {\bibinfo {author} {\bibfnamefont {P.}~\bibnamefont
  {Coleman}}\ and\ \bibinfo {author} {\bibfnamefont {N.}~\bibnamefont
  {Andrei}},\ }\href@noop {} {\bibfield  {journal} {\bibinfo  {journal} {Jour.
  Phys.}\ }\textbf {\bibinfo {volume} {C19}},\ \bibinfo {pages} {3211}
  (\bibinfo {year} {1986})}\BibitemShut {NoStop}%
\bibitem [{\citenamefont {Coleman}(1987)}]{PhysRevB.35.5072}%
  \BibitemOpen
  \bibfield  {author} {\bibinfo {author} {\bibfnamefont {P.}~\bibnamefont
  {Coleman}},\ }\href {\doibase 10.1103/PhysRevB.35.5072} {\bibfield  {journal}
  {\bibinfo  {journal} {Phys. Rev.}\ }\textbf {\bibinfo {volume} {B35}},\
  \bibinfo {pages} {5072} (\bibinfo {year} {1987})}\BibitemShut {NoStop}%
\bibitem [{\citenamefont {Senthil}\ \emph {et~al.}(2003)\citenamefont
  {Senthil}, \citenamefont {Sachdev},\ and\ \citenamefont
  {Vojta}}]{2003PhRvL..90u6403S}%
  \BibitemOpen
  \bibfield  {author} {\bibinfo {author} {\bibfnamefont {T.}~\bibnamefont
  {Senthil}}, \bibinfo {author} {\bibfnamefont {S.}~\bibnamefont {Sachdev}}, \
  and\ \bibinfo {author} {\bibfnamefont {M.}~\bibnamefont {Vojta}},\ }\href
  {\doibase 10.1103/PhysRevLett.90.216403} {\bibfield  {journal} {\bibinfo
  {journal} {Phys. Rev. Lett.}\ }\textbf {\bibinfo {volume} {90}},\ \bibinfo
  {eid} {216403} (\bibinfo {year} {2003})},\ \Eprint
  {http://arxiv.org/abs/{arXiv:cond-mat/0209144}} {{arXiv:cond-mat/0209144}}
  \BibitemShut {NoStop}%
\bibitem [{\citenamefont {Senthil}\ \emph {et~al.}(2004)\citenamefont
  {Senthil}, \citenamefont {Vojta},\ and\ \citenamefont
  {Sachdev}}]{2004PhRvB..69c5111S}%
  \BibitemOpen
  \bibfield  {author} {\bibinfo {author} {\bibfnamefont {T.}~\bibnamefont
  {Senthil}}, \bibinfo {author} {\bibfnamefont {M.}~\bibnamefont {Vojta}}, \
  and\ \bibinfo {author} {\bibfnamefont {S.}~\bibnamefont {Sachdev}},\ }\href
  {\doibase 10.1103/PhysRevB.69.035111} {\bibfield  {journal} {\bibinfo
  {journal} {Phys. Rev.}\ }\textbf {\bibinfo {volume} {B69}},\ \bibinfo {eid}
  {035111} (\bibinfo {year} {2004})},\ \Eprint
  {http://arxiv.org/abs/{arxiv:cond-mat/0305193}} {{arxiv:cond-mat/0305193}}
  \BibitemShut {NoStop}%
\bibitem [{\citenamefont {Fano}(1961)}]{PhysRev.124.1866}%
  \BibitemOpen
  \bibfield  {author} {\bibinfo {author} {\bibfnamefont {U.}~\bibnamefont
  {Fano}},\ }\href {\doibase 10.1103/PhysRev.124.1866} {\bibfield  {journal}
  {\bibinfo  {journal} {Phys. Rev.}\ }\textbf {\bibinfo {volume} {124}},\
  \bibinfo {pages} {1866} (\bibinfo {year} {1961})}\BibitemShut {NoStop}%
\bibitem [{\citenamefont {Miroshnichenko}\ \emph {et~al.}(2010)\citenamefont
  {Miroshnichenko}, \citenamefont {Flach},\ and\ \citenamefont
  {Kivshar}}]{RevModPhys.82.2257}%
  \BibitemOpen
  \bibfield  {author} {\bibinfo {author} {\bibfnamefont {A.~E.}\ \bibnamefont
  {Miroshnichenko}}, \bibinfo {author} {\bibfnamefont {S.}~\bibnamefont
  {Flach}}, \ and\ \bibinfo {author} {\bibfnamefont {Y.~S.}\ \bibnamefont
  {Kivshar}},\ }\href {\doibase 10.1103/RevModPhys.82.2257} {\bibfield
  {journal} {\bibinfo  {journal} {Rev. Mod. Phys.}\ }\textbf {\bibinfo {volume}
  {82}},\ \bibinfo {pages} {2257} (\bibinfo {year} {2010})}\BibitemShut
  {NoStop}%
\bibitem [{\citenamefont {Madhavan}\ \emph {et~al.}(1998)\citenamefont
  {Madhavan}, \citenamefont {Chen}, \citenamefont {Jamneala}, \citenamefont
  {Crommie},\ and\ \citenamefont {Wingreen}}]{Madhavan567}%
  \BibitemOpen
  \bibfield  {author} {\bibinfo {author} {\bibfnamefont {V.}~\bibnamefont
  {Madhavan}}, \bibinfo {author} {\bibfnamefont {W.}~\bibnamefont {Chen}},
  \bibinfo {author} {\bibfnamefont {T.}~\bibnamefont {Jamneala}}, \bibinfo
  {author} {\bibfnamefont {M.~F.}\ \bibnamefont {Crommie}}, \ and\ \bibinfo
  {author} {\bibfnamefont {N.~S.}\ \bibnamefont {Wingreen}},\ }\href {\doibase
  10.1126/science.280.5363.567} {\bibfield  {journal} {\bibinfo  {journal}
  {Science}\ }\textbf {\bibinfo {volume} {280}},\ \bibinfo {pages} {567}
  (\bibinfo {year} {1998})}\BibitemShut {NoStop}%
\bibitem [{\citenamefont {Madhavan}\ \emph {et~al.}(2001)\citenamefont
  {Madhavan}, \citenamefont {Chen}, \citenamefont {Jamneala}, \citenamefont
  {Crommie},\ and\ \citenamefont {Wingreen}}]{PhysRevB.64.165412}%
  \BibitemOpen
  \bibfield  {author} {\bibinfo {author} {\bibfnamefont {V.}~\bibnamefont
  {Madhavan}}, \bibinfo {author} {\bibfnamefont {W.}~\bibnamefont {Chen}},
  \bibinfo {author} {\bibfnamefont {T.}~\bibnamefont {Jamneala}}, \bibinfo
  {author} {\bibfnamefont {M.~F.}\ \bibnamefont {Crommie}}, \ and\ \bibinfo
  {author} {\bibfnamefont {N.~S.}\ \bibnamefont {Wingreen}},\ }\href {\doibase
  10.1103/PhysRevB.64.165412} {\bibfield  {journal} {\bibinfo  {journal} {Phys.
  Rev. B}\ }\textbf {\bibinfo {volume} {64}},\ \bibinfo {pages} {165412}
  (\bibinfo {year} {2001})}\BibitemShut {NoStop}%
\bibitem [{\citenamefont {{G{\"o}res}}\ \emph {et~al.}(2000)\citenamefont
  {{G{\"o}res}}, \citenamefont {{Goldhaber-Gordon}}, \citenamefont
  {{Heemeyer}}, \citenamefont {{Kastner}}, \citenamefont {{Shtrikman}},
  \citenamefont {{Mahalu}},\ and\ \citenamefont
  {{Meirav}}}]{2000PhRvB..62.2188G}%
  \BibitemOpen
  \bibfield  {author} {\bibinfo {author} {\bibfnamefont {J.}~\bibnamefont
  {{G{\"o}res}}}, \bibinfo {author} {\bibfnamefont {D.}~\bibnamefont
  {{Goldhaber-Gordon}}}, \bibinfo {author} {\bibfnamefont {S.}~\bibnamefont
  {{Heemeyer}}}, \bibinfo {author} {\bibfnamefont {M.~A.}\ \bibnamefont
  {{Kastner}}}, \bibinfo {author} {\bibfnamefont {H.}~\bibnamefont
  {{Shtrikman}}}, \bibinfo {author} {\bibfnamefont {D.}~\bibnamefont
  {{Mahalu}}}, \ and\ \bibinfo {author} {\bibfnamefont {U.}~\bibnamefont
  {{Meirav}}},\ }\href {\doibase 10.1103/PhysRevB.62.2188} {\bibfield
  {journal} {\bibinfo  {journal} {Phys. Rev. B}\ }\textbf {\bibinfo {volume}
  {62}},\ \bibinfo {pages} {2188} (\bibinfo {year} {2000})},\ \Eprint
  {http://arxiv.org/abs/cond-mat/9912419} {cond-mat/9912419} \BibitemShut
  {NoStop}%
\bibitem [{\citenamefont {Parcollet}\ \emph {et~al.}(1998)\citenamefont
  {Parcollet}, \citenamefont {Georges}, \citenamefont {Kotliar},\ and\
  \citenamefont {Sengupta}}]{1998PhRvB..58.3794P}%
  \BibitemOpen
  \bibfield  {author} {\bibinfo {author} {\bibfnamefont {O.}~\bibnamefont
  {Parcollet}}, \bibinfo {author} {\bibfnamefont {A.}~\bibnamefont {Georges}},
  \bibinfo {author} {\bibfnamefont {G.}~\bibnamefont {Kotliar}}, \ and\
  \bibinfo {author} {\bibfnamefont {A.}~\bibnamefont {Sengupta}},\ }\href
  {\doibase 10.1103/PhysRevB.58.3794} {\bibfield  {journal} {\bibinfo
  {journal} {Phys. Rev.}\ }\textbf {\bibinfo {volume} {B58}},\ \bibinfo {pages}
  {3794} (\bibinfo {year} {1998})},\ \Eprint
  {http://arxiv.org/abs/{arXiv:cond-mat/9711192}} {{arXiv:cond-mat/9711192}}
  \BibitemShut {NoStop}%
\bibitem [{\citenamefont {Faulkner}\ \emph
  {et~al.}(2011{\natexlab{a}})\citenamefont {Faulkner}, \citenamefont {Liu},
  \citenamefont {McGreevy},\ and\ \citenamefont {Vegh}}]{Faulkner:2009wj}%
  \BibitemOpen
  \bibfield  {author} {\bibinfo {author} {\bibfnamefont {T.}~\bibnamefont
  {Faulkner}}, \bibinfo {author} {\bibfnamefont {H.}~\bibnamefont {Liu}},
  \bibinfo {author} {\bibfnamefont {J.}~\bibnamefont {McGreevy}}, \ and\
  \bibinfo {author} {\bibfnamefont {D.}~\bibnamefont {Vegh}},\ }\href {\doibase
  10.1103/PhysRevD.83.125002} {\bibfield  {journal} {\bibinfo  {journal} {Phys.
  Rev.}\ }\textbf {\bibinfo {volume} {D83}},\ \bibinfo {pages} {125002}
  (\bibinfo {year} {2011}{\natexlab{a}})},\ \Eprint
  {http://arxiv.org/abs/0907.2694} {arXiv:0907.2694 [hep-th]} \BibitemShut
  {NoStop}%
\bibitem [{\citenamefont {Faulkner}\ \emph
  {et~al.}(2011{\natexlab{b}})\citenamefont {Faulkner}, \citenamefont {Iqbal},
  \citenamefont {Liu}, \citenamefont {McGreevy},\ and\ \citenamefont
  {Vegh}}]{Faulkner:2011tm}%
  \BibitemOpen
  \bibfield  {author} {\bibinfo {author} {\bibfnamefont {T.}~\bibnamefont
  {Faulkner}}, \bibinfo {author} {\bibfnamefont {N.}~\bibnamefont {Iqbal}},
  \bibinfo {author} {\bibfnamefont {H.}~\bibnamefont {Liu}}, \bibinfo {author}
  {\bibfnamefont {J.}~\bibnamefont {McGreevy}}, \ and\ \bibinfo {author}
  {\bibfnamefont {D.}~\bibnamefont {Vegh}},\ }\href {\doibase
  10.1098/rsta.2010.0354} {\bibfield  {journal} {\bibinfo  {journal} {Phil.
  Trans. Roy. Soc.}\ }\textbf {\bibinfo {volume} {A 369}},\ \bibinfo {pages}
  {1640} (\bibinfo {year} {2011}{\natexlab{b}})},\ \Eprint
  {http://arxiv.org/abs/1101.0597} {arXiv:1101.0597 [hep-th]} \BibitemShut
  {NoStop}%
\bibitem [{\citenamefont {Sachdev}(2015)}]{Sachdev:2015efa}%
  \BibitemOpen
  \bibfield  {author} {\bibinfo {author} {\bibfnamefont {S.}~\bibnamefont
  {Sachdev}},\ }\href {\doibase 10.1103/PhysRevX.5.041025} {\bibfield
  {journal} {\bibinfo  {journal} {Phys. Rev.}\ }\textbf {\bibinfo {volume}
  {X5}},\ \bibinfo {pages} {041025} (\bibinfo {year} {2015})},\ \Eprint
  {http://arxiv.org/abs/1506.05111} {arXiv:1506.05111 [hep-th]} \BibitemShut
  {NoStop}%
\bibitem [{\citenamefont {Sachdev}\ and\ \citenamefont
  {Ye}(1993)}]{Sachdev:1992fk}%
  \BibitemOpen
  \bibfield  {author} {\bibinfo {author} {\bibfnamefont {S.}~\bibnamefont
  {Sachdev}}\ and\ \bibinfo {author} {\bibfnamefont {J.-W.}\ \bibnamefont
  {Ye}},\ }\href {\doibase 10.1103/PhysRevLett.70.3339} {\bibfield  {journal}
  {\bibinfo  {journal} {Phys. Rev. Lett.}\ }\textbf {\bibinfo {volume} {70}},\
  \bibinfo {pages} {3339} (\bibinfo {year} {1993})},\ \Eprint
  {http://arxiv.org/abs/cond-mat/9212030} {arXiv:cond-mat/9212030 [cond-mat]}
  \BibitemShut {NoStop}%
\bibitem [{\citenamefont {Kitaev}(2015)}]{kitaev}%
  \BibitemOpen
  \bibfield  {author} {\bibinfo {author} {\bibfnamefont {A.}~\bibnamefont
  {Kitaev}},\ }\href {http://online.kitp.ucsb.edu/online/entangled15/}
  {\enquote {\bibinfo {title} {A simple model of quantum holography},}\ }
  (\bibinfo {year} {2015}),\ \bibinfo {note} {talks for the KITP Strings
  seminar and Entanglement 2015 program, Feb. 12, Apr. 7, and May 27,
  2015}\BibitemShut {NoStop}%
\bibitem [{\citenamefont {Polchinski}\ and\ \citenamefont
  {Rosenhaus}(2016)}]{Polchinski:2016xgd}%
  \BibitemOpen
  \bibfield  {author} {\bibinfo {author} {\bibfnamefont {J.}~\bibnamefont
  {Polchinski}}\ and\ \bibinfo {author} {\bibfnamefont {V.}~\bibnamefont
  {Rosenhaus}},\ }\href {\doibase 10.1007/JHEP04(2016)001} {\bibfield
  {journal} {\bibinfo  {journal} {JHEP}\ }\textbf {\bibinfo {volume} {04}},\
  \bibinfo {pages} {001} (\bibinfo {year} {2016})},\ \Eprint
  {http://arxiv.org/abs/1601.06768} {arXiv:1601.06768 [hep-th]} \BibitemShut
  {NoStop}%
\bibitem [{\citenamefont {Jevicki}\ \emph {et~al.}(2016)\citenamefont
  {Jevicki}, \citenamefont {Suzuki},\ and\ \citenamefont
  {Yoon}}]{Jevicki:2016bwu}%
  \BibitemOpen
  \bibfield  {author} {\bibinfo {author} {\bibfnamefont {A.}~\bibnamefont
  {Jevicki}}, \bibinfo {author} {\bibfnamefont {K.}~\bibnamefont {Suzuki}}, \
  and\ \bibinfo {author} {\bibfnamefont {J.}~\bibnamefont {Yoon}},\ }\href
  {\doibase 10.1007/JHEP07(2016)007} {\bibfield  {journal} {\bibinfo  {journal}
  {JHEP}\ }\textbf {\bibinfo {volume} {07}},\ \bibinfo {pages} {007} (\bibinfo
  {year} {2016})},\ \Eprint {http://arxiv.org/abs/1603.06246} {arXiv:1603.06246
  [hep-th]} \BibitemShut {NoStop}%
\bibitem [{\citenamefont {Maldacena}\ and\ \citenamefont
  {Stanford}(2016)}]{Maldacena:2016hyu}%
  \BibitemOpen
  \bibfield  {author} {\bibinfo {author} {\bibfnamefont {J.}~\bibnamefont
  {Maldacena}}\ and\ \bibinfo {author} {\bibfnamefont {D.}~\bibnamefont
  {Stanford}},\ }\href {\doibase 10.1103/PhysRevD.94.106002} {\bibfield
  {journal} {\bibinfo  {journal} {Phys. Rev.}\ }\textbf {\bibinfo {volume}
  {D94}},\ \bibinfo {pages} {106002} (\bibinfo {year} {2016})},\ \Eprint
  {http://arxiv.org/abs/1604.07818} {arXiv:1604.07818 [hep-th]} \BibitemShut
  {NoStop}%
\bibitem [{\citenamefont {Jevicki}\ and\ \citenamefont
  {Suzuki}(2016)}]{Jevicki:2016ito}%
  \BibitemOpen
  \bibfield  {author} {\bibinfo {author} {\bibfnamefont {A.}~\bibnamefont
  {Jevicki}}\ and\ \bibinfo {author} {\bibfnamefont {K.}~\bibnamefont
  {Suzuki}},\ }\href {\doibase 10.1007/JHEP11(2016)046} {\bibfield  {journal}
  {\bibinfo  {journal} {JHEP}\ }\textbf {\bibinfo {volume} {11}},\ \bibinfo
  {pages} {046} (\bibinfo {year} {2016})},\ \Eprint
  {http://arxiv.org/abs/1608.07567} {arXiv:1608.07567 [hep-th]} \BibitemShut
  {NoStop}%
\bibitem [{\citenamefont {Witten}(2016)}]{Witten:2016iux}%
  \BibitemOpen
  \bibfield  {author} {\bibinfo {author} {\bibfnamefont {E.}~\bibnamefont
  {Witten}},\ }\href@noop {} {\  (\bibinfo {year} {2016})},\ \Eprint
  {http://arxiv.org/abs/1610.09758} {arXiv:1610.09758 [hep-th]} \BibitemShut
  {NoStop}%
\bibitem [{\citenamefont {Erdmenger}\ \emph {et~al.}(2017)\citenamefont
  {Erdmenger}, \citenamefont {Hoyos}, \citenamefont {O'Bannon}, \citenamefont
  {Papadimitriou}, \citenamefont {Probst},\ and\ \citenamefont
  {Wu}}]{Erdmenger:2016jjg}%
  \BibitemOpen
  \bibfield  {author} {\bibinfo {author} {\bibfnamefont {J.}~\bibnamefont
  {Erdmenger}}, \bibinfo {author} {\bibfnamefont {C.}~\bibnamefont {Hoyos}},
  \bibinfo {author} {\bibfnamefont {A.}~\bibnamefont {O'Bannon}}, \bibinfo
  {author} {\bibfnamefont {I.}~\bibnamefont {Papadimitriou}}, \bibinfo {author}
  {\bibfnamefont {J.}~\bibnamefont {Probst}}, \ and\ \bibinfo {author}
  {\bibfnamefont {J.~M.~S.}\ \bibnamefont {Wu}},\ }\href {\doibase
  10.1007/JHEP03(2017)039} {\bibfield  {journal} {\bibinfo  {journal} {JHEP}\
  }\textbf {\bibinfo {volume} {03}},\ \bibinfo {pages} {039} (\bibinfo {year}
  {2017})},\ \Eprint {http://arxiv.org/abs/1612.02005} {arXiv:1612.02005
  [hep-th]} \BibitemShut {NoStop}%
\bibitem [{\citenamefont {Affleck}(1995)}]{Affleck:1995ge}%
  \BibitemOpen
  \bibfield  {author} {\bibinfo {author} {\bibfnamefont {I.}~\bibnamefont
  {Affleck}},\ }\href@noop {} {\bibfield  {journal} {\bibinfo  {journal} {Acta
  Phys. Polon.}\ }\textbf {\bibinfo {volume} {B26}},\ \bibinfo {pages} {1869}
  (\bibinfo {year} {1995})},\ \Eprint {http://arxiv.org/abs/cond-mat/9512099}
  {arXiv:cond-mat/9512099} \BibitemShut {NoStop}%
\bibitem [{\citenamefont {{Bickers, N.}}(1987)}]{RevModPhys.59.845}%
  \BibitemOpen
  \bibfield  {author} {\bibinfo {author} {\bibnamefont {{Bickers, N.}}},\
  }\href {\doibase 10.1103/RevModPhys.59.845} {\bibfield  {journal} {\bibinfo
  {journal} {Rev. Mod. Phys.}\ }\textbf {\bibinfo {volume} {59}},\ \bibinfo
  {pages} {845} (\bibinfo {year} {1987})}\BibitemShut {NoStop}%
\bibitem [{\citenamefont {Coleman}(2007)}]{2006cond.mat.12006C}%
  \BibitemOpen
  \bibfield  {author} {\bibinfo {author} {\bibfnamefont {P.}~\bibnamefont
  {Coleman}},\ }in\ \href@noop {} {\emph {\bibinfo {booktitle} {Handbook of
  Magnetism and Advanced Magnetic Materials: Fundamentals and Theory}}},\
  Vol.~\bibinfo {volume} {1},\ \bibinfo {editor} {edited by\ \bibinfo {editor}
  {\bibnamefont {Kronmuller}}\ and\ \bibinfo {editor} {\bibnamefont {Parkin}}}\
  (\bibinfo  {publisher} {John Wiley and Sons},\ \bibinfo {year} {2007})\ pp.\
  \bibinfo {pages} {95--148},\ \Eprint
  {http://arxiv.org/abs/[arxiv:cond-mat/0612006]} {[arxiv:cond-mat/0612006]}
  \BibitemShut {NoStop}%
\bibitem [{\citenamefont {Aharony}\ \emph {et~al.}(2000)\citenamefont
  {Aharony}, \citenamefont {Gubser}, \citenamefont {Maldacena}, \citenamefont
  {Ooguri},\ and\ \citenamefont {Oz}}]{Aharony:1999ti}%
  \BibitemOpen
  \bibfield  {author} {\bibinfo {author} {\bibfnamefont {O.}~\bibnamefont
  {Aharony}}, \bibinfo {author} {\bibfnamefont {S.~S.}\ \bibnamefont {Gubser}},
  \bibinfo {author} {\bibfnamefont {J.~M.}\ \bibnamefont {Maldacena}}, \bibinfo
  {author} {\bibfnamefont {H.}~\bibnamefont {Ooguri}}, \ and\ \bibinfo {author}
  {\bibfnamefont {Y.}~\bibnamefont {Oz}},\ }\href {\doibase
  10.1016/S0370-1573(99)00083-6} {\bibfield  {journal} {\bibinfo  {journal}
  {Phys. Rept.}\ }\textbf {\bibinfo {volume} {323}},\ \bibinfo {pages} {183}
  (\bibinfo {year} {2000})},\ \Eprint {http://arxiv.org/abs/hep-th/9905111}
  {arXiv:hep-th/9905111 [hep-th]} \BibitemShut {NoStop}%
\bibitem [{\citenamefont {Witten}(2001)}]{Witten:2001ua}%
  \BibitemOpen
  \bibfield  {author} {\bibinfo {author} {\bibfnamefont {E.}~\bibnamefont
  {Witten}},\ }\href@noop {} {\  (\bibinfo {year} {2001})},\ \Eprint
  {http://arxiv.org/abs/hep-th/0112258} {arXiv:hep-th/0112258 [hep-th]}
  \BibitemShut {NoStop}%
\bibitem [{\citenamefont {Papadimitriou}(2007)}]{Papadimitriou:2007sj}%
  \BibitemOpen
  \bibfield  {author} {\bibinfo {author} {\bibfnamefont {I.}~\bibnamefont
  {Papadimitriou}},\ }\href {\doibase 10.1088/1126-6708/2007/05/075} {\bibfield
   {journal} {\bibinfo  {journal} {JHEP}\ }\textbf {\bibinfo {volume} {0705}},\
  \bibinfo {pages} {075} (\bibinfo {year} {2007})},\ \Eprint
  {http://arxiv.org/abs/hep-th/0703152} {arXiv:hep-th/0703152 [hep-th]}
  \BibitemShut {NoStop}%
\bibitem [{\citenamefont {Son}\ and\ \citenamefont
  {Starinets}(2002)}]{Son:2002sd}%
  \BibitemOpen
  \bibfield  {author} {\bibinfo {author} {\bibfnamefont {D.~T.}\ \bibnamefont
  {Son}}\ and\ \bibinfo {author} {\bibfnamefont {A.~O.}\ \bibnamefont
  {Starinets}},\ }\href@noop {} {\bibfield  {journal} {\bibinfo  {journal}
  {JHEP}\ }\textbf {\bibinfo {volume} {09}},\ \bibinfo {pages} {042} (\bibinfo
  {year} {2002})},\ \Eprint {http://arxiv.org/abs/hep-th/0205051}
  {arXiv:hep-th/0205051} \BibitemShut {NoStop}%
\bibitem [{\citenamefont {Kovtun}\ and\ \citenamefont
  {Starinets}(2005)}]{Kovtun:2005ev}%
  \BibitemOpen
  \bibfield  {author} {\bibinfo {author} {\bibfnamefont {P.~K.}\ \bibnamefont
  {Kovtun}}\ and\ \bibinfo {author} {\bibfnamefont {A.~O.}\ \bibnamefont
  {Starinets}},\ }\href {\doibase 10.1103/PhysRevD.72.086009} {\bibfield
  {journal} {\bibinfo  {journal} {Phys. Rev.}\ }\textbf {\bibinfo {volume}
  {D72}},\ \bibinfo {pages} {086009} (\bibinfo {year} {2005})},\ \Eprint
  {http://arxiv.org/abs/hep-th/0506184} {arXiv:hep-th/0506184} \BibitemShut
  {NoStop}%
\end{thebibliography}%
\end{document}